\documentstyle[preprint,prb,aps]{revtex}
\begin{document}
\title{Density-functional theory of quantum wires and dots in a
strong magnetic
field}
\author{T.H. Stoof and Gerrit E.W. Bauer}
\address{
Department of Applied Physics and Delft Institute for
Microelectronics and 
Submicrontechnology,\\ Delft University of Technology, Lorentzweg 1,
2628 CJ
Delft, The Netherlands}
\date{\today}
\maketitle
\begin{abstract}
We study the competition between the exchange and the direct
Coulomb interaction
near the edge of a two-dimensional electron gas in a strong
magnetic field
using density-functional theory in a local approximation for the
exchange-energy functional. Exchange is shown to play a significant
role in
reducing the
spatial extent of the compressible edge channel regions obtained
from an
electrostatic description. The transition from the incompressible
edge
channels of the Hartree-Fock picture to the broad, compressible
strips
predicted by electrostatics occurs within a narrow and
experimentally 
accessible range of confinement strengths.
    \\
Submitted to Phys. Rev. B.
\end{abstract}


\section{Introduction}
\label{sec:introduction}

The concept of current-carrying edge channels accounts for the
magnetotransport 
properties of a two-dimensional electron gas (2DEG) in a high
magnetic
field, both in the integer\cite{halperin} and
fractional\cite{been,macdonald,wen} quantum Hall regime.
Although the
initial theoretical studies have used a noninteracting picture of
edge channels,
a considerable effort has recently been devoted to understand the
effects of
electron-electron interactions in the {\em integer} quantum Hall
regime.\cite{been,macdonald,wen,mej,chklovskii,cms,dempsey,chamon}

At present there are two incompatible pictures for the electronic
ground state of edge channels in the integer quantum Hall regime. In
the Hartree-Fock approximation the ground state wave function is a
single Slater
determinant, which corresponds to occupation numbers zero and
one.\cite{dempsey}
In this description the edge state is incompressible and the
electron density 
drops rapidly at the edges, on a length scale which is typically
of the order of the magnetic length. Although the Hartree-Fock
approximation
is widely used for the description of quantum dots and
wires,\cite{mej,dempsey,chamon} it has been challenged on the
grounds that is
does not take the global electrostatics into account
properly.\cite{chklovskii,cms}
In the electrostatic description given in
Ref.~\onlinecite{chklovskii} a more
gradual variation of the electron density at the edge is found
energetically
favorable. Here the typical length scale for the density drop at the
edge is of the order of the depletion length, which is much larger
than the
magnetic length. In this picture the electrons at the edge can
screen the
electrostatic confinement potential by a slow density variation and
the edge
states are therefore compressible. However, this description
completely neglects the exchange interaction. 

A transition from incompressible to compressible edge states was
already 
qualitatively discussed in Ref.~\onlinecite{dempsey}, where a
spontaneous
transition from an unpolarized ({\it i.e.} equal occupation for both
spin levels) to a polarized (different occupation for different spin
levels) 
Hartree-Fock ground state was found for a
critical confinement strength. It was speculated that a development
to the electrostatic regime takes place as the confinement strength
is
decreased.
More recently the transition between smoothly and abruptly varying
density
distributions has been studied by Chamon and Wen.\cite{chamon} On
the basis
of few particle exact solutions they predict formation of
compressible edge
states when the strength of the confinement potential is reduced
beyond a
certain point. Their approach is difficult to extend to quantum
wires or dots
with a large number of electrons.
 
In the present work we interpolate between the different regimes of
applicability of both Hartree-Fock and electrostatic pictures by
extending the
Thomas-Fermi approach used in
Ref.\onlinecite{chklovskii} to a Thomas-Fermi-{\em Dirac} like
treatment of the exchange effects in strong magnetic fields. First
results
of the present approach have been published in
Ref.~\onlinecite{stoof}. Very
recently Ferconi and Vignale\cite{ferconi} have studied the ground
state
energies and densities of a quantum dot in an arbitrary magnetic
field, taking
into account exchange-correlation effects by employing a Kohn-Sham
scheme of
current-density functional theory.\cite{vignale} For small quantum
dots
(2 or 3 electrons) their method yields an accuracy better than 3\%
when
compared with exact results. For a larger number of electrons and
high magnetic
fields we find that the Kohn-Sham formalism encounters serious
problems.

In Sec.~\ref{sec:DFT} we describe our implementation of
density-functional
theory in strong magnetic fields as applied to quantum wires and
derive
expressions for the density profile and single-particle potentials
of the
ground state. In Sec.~\ref{sec:results} we present
results of the numerical calculations and we investigate both the
accuracy of
our theory and the relation with possible experiments. Two
experimentally
relevant applications are studied in Sec.~\ref{sec:applications};
the influence of a plane of constant potential parallel to the 2DEG on
 the
electronic ground state and the ground state properties of a quantum
dot in a
strong magnetic field. Section~\ref{sec:conclusions} summarizes our
conclusions.

\section{Density-functional theory in strong magnetic fields}
\label{sec:DFT}

\subsection{The system}
\label{sec:system}

Let us consider first a quantum wire of the strictly two dimensional
electron
gas along the $y$~axis in the $xy$~plane (Fig.~\ref{fig:system}).
Perpendicular to the plane a strong uniform magnetic field is applied.
Also present is a uniform positive background charge which ensures
global
neutrality. We use periodic boundary conditions in the $y$-direction
and adopt
the Landau gauge so that ${\bf A}_{0}=B_{0} x \hat{\bf y}$ and
${\bf B}_{0}=B_{0} \hat{\bf z}$. In the high magnetic field limit
considered
here only the lowest (spin-up and spin-down) Landau levels are
occupied.

The wire is confined in the $x$~direction by a parabolic confinement
potential:
\begin{equation}
V_{c}(x)\;=\; \alpha \; \frac{e}{4 \pi \epsilon l_{B}} \; x^{2}, 
\label{confinement}
\end{equation}
where $\epsilon = \epsilon_{0} \epsilon_{r}$ is the dielectric
constant and 
$\alpha$ is a dimensionless parameter. 
We will ignore here the difference between the 'bare' magnetic length
$l_{B}=\sqrt{\hbar / e B_{0}}$ and the magnetic length $l$ which is
renormalized
by the parabolic confinement potential:
\begin{equation}
l = l_{B} \sqrt{ \frac{\omega_{c}^{2}}{\omega_{c}^{2} +
\omega_{0}^{2}} },
\label{renmaglen}
\end{equation}
where $\omega_{c}=eB_{0}/m^{*}$ is the cyclotron frequency with
$m^{*}$ the
effective electron mass and $\omega_{0}$ characterizes the parabolic
confinement
potential $\frac{1}{2} m^{*} \omega_{0}^{2} x^{2}$. Since we treat the
problem numerically,
the assumption of a parabolic form for the confinement is not
essential: It
is chosen here because it is widely used in the literature. All
lengths will be
given in units of $l_{B}$ throughout the paper unless otherwise
indicated.

\subsection{Theory}
\label{sec:theory}

We use density-functional theory in strong magnetic fields to find the
electronic
ground state of the system.\cite{vignale} Without a magnetic field,
the total
energy of the system would be a functional of the electron density
only.
However, a magnetic field, apart from giving rise to Zeeman splitting,
 causes 
orbital currents to flow in the electron gas, even when the system is
in thermodynamic equilibrium. As a consequence, the total energy
is now a functional of the density
$n({\bf r})$, the spin density ${\bf s}({\bf r})$ and the paramagnetic
 current
density ${\bf j}_{p}({\bf r})$.\cite{vignale}

We do not take into account correlation effects, which means that we
operate
strictly in the integer
quantum Hall regime, since correlation is responsible for the energy
gaps
that cause the fractional quantization. Due to the fact that the
correlation
part of the energy is disregarded and exchange does not depend on the
current
density the total energy functional depends only
on the total and the spin density or equivalently, since the
spin-quantization
axis is well defined by the strong magnetic field, on the density of
spin up
and spin down electrons.

In the Kohn-Sham scheme of density-functional theory the ground state
densities
$n^{\sigma}$ of the interacting electron system are expressed in
terms
of a set of $M^{\sigma}$ Kohn-Sham orbitals
$\phi^{\sigma}_{i}({\bf r})$:
\begin{equation}
n^{\sigma}({\bf r}) = \sum_{i=1}^{M^{\sigma}} {\left| 
\phi^{\sigma}_{i}({\bf r}) \right|}^{2},
\label{ksdensity}
\end{equation}
where $\sigma=\uparrow,\downarrow$ denotes up or down spin
respectively and
$M^{\sigma}$ is the number of electrons with spin $\sigma$. These
orbitals
satisfy the Kohn-Sham equations:
\begin{equation}
\left\{ \frac{-{\hbar}^{2}}{2 m^{*}} {\nabla}^{2} +
V^{\sigma}_{\mbox{\small eff}}({\bf r}) \right\}
\phi^{\sigma}_{i}({\bf r}) =
\epsilon^{\sigma}_{i} \phi^{\sigma}_{i}({\bf r}),
\label{ksequation}
\end{equation}
with $V^{\sigma}_{\mbox{\small eff}}({\bf r})$ the effective
one-particle
potential:
\begin{equation}
V^{\sigma}_{\mbox{\small eff}}({\bf r}) = V_{c}({\bf r}) +
V_{H}([n];{\bf r}) +
V^{\sigma}_{x}([n^{\uparrow},n^{\downarrow}]
;{\bf r}),
\label{veff}
\end{equation}
where $V_{H}$ is the Hartree potential and $V^{\sigma}_{x}$ the
exchange potential. Here we have disregarded the
exchange-correlation vector potential ${\bf A}_{xc}$ which gives only
a very
small contribution to the total energy.\cite{ferconi}
The Kohn-Sham ground state wave function is a Slater determinant of
the
$M^{\sigma}$ lowest Kohn-Sham orbitals:
\begin{equation}
\Phi^{\sigma}({\bf r}_{1},{\bf r}_{2},...,{\bf r}_{M^{\sigma}}) =
\frac{1}{\sqrt{M^{\sigma}}} \mbox{det}
\left\{ \phi^{\sigma}_{j}({\bf r}_{i})
\right\}.
\label{ksground}
\end{equation}

It can be shown (see Sec.~\ref{sec:exp}) that in the limit of high
magnetic
fields the mixing between Landau level wave functions vanishes.
Consider,
{\it e.g.}, the case of a single occupied, spin resolved Landau level.
 For large
magnetic fields the Kohn-Sham ground state wave function,
Eq.~(\ref{ksground}),
reduces to a Slater determinant of lowest Landau level
wave functions with integer filling. As a consequence the scheme
cannot describe fractional filling in the extreme quantum limit we are
interested in. The reason for the failure of this procedure is the
fact that
the crucial assumption of the Kohn-Sham scheme, namely that the
interacting
$v$-representable densities are also non-interacting
$v$-representable,\cite{buukske} breaks down in the high magnetic
field limit.

It is not possible to extend the Kohn-Sham scheme to fractional
filling since
in that case the effective potential,
$V^{\sigma}_{\mbox{\small eff}}({\bf r})$,
is no longer a unique functional of the total and spin density.
However, at
sufficiently high magnetic fields the basis wave functions are known
to be
just Landau level wave functions. We may then use a variational scheme
 which
allows fractional filling and in which
$E[n^{\uparrow},n^{\downarrow}]$ is a
unique functional of the total and spin density.

The equilibrium density distribution
$n(x)=n^{\uparrow}(x)+n^{\downarrow}(x)$
which minimizes the total energy $E$ of the system can be found by
solving:
\begin{equation}
\frac{\delta E[n^{\uparrow},n^{\downarrow}]}{\delta n^{\uparrow}(x)}=
\frac{\delta E[n^{\uparrow},n^{\downarrow}]}{\delta n^{\downarrow}(x)}
=\mu,
\label{chempot}
\end{equation}
where $\mu$ is the chemical potential and the total energy of the
quantum wire is given by:
\begin{equation}
E[n^{\uparrow},n^{\downarrow}] = T[n] + E_{c}[n] + E_{H}[n] +
E_{Z}[n^{\uparrow},n^{\downarrow}] +
E_{x}[n^{\uparrow},n^{\downarrow}].
\label{totenergy}
\end{equation}

The total energy, Eq.~(\ref{totenergy}), consists of five
contributions:
The first term on the right
hand side is the kinetic energy which we define as a functional of
the total
density:
\begin{equation}
T[n] = \sum_{N \sigma} \int_{-\infty}^{\infty}
dx \; n_{N}^{\sigma}(x)\;
(N+\frac{1}{2}) \hbar \omega_{c},
\label{kinen}
\end{equation}
where $N$ labels the Landau levels.
The densities $n_{N}^{\sigma}(x)$ are the partial densities for a
given spin
direction $\sigma$ and Landau level $N$. The total electron density is
 found by
summing over all occupied Landau levels and over spin directions:
$n(x)=\sum_{N \sigma} n^{\sigma}_{N}(x)$ and is for sufficiently high
magnetic fields given by:
\begin{equation}
n(x)= \sum_{\sigma N X} \; \nu_{N X}^{\sigma} \;
{\left| \psi_{N X}(x,y)
\right|}^{2},
\label{genconv}
\end{equation}
where $X \equiv k_{y} l_{B}^{2}$ is the quantum number of an electron
with
momentum $k_{y}$, $\psi_{N X}(x,y) \sim \phi_{N X}(x)
\exp{(iXy/l_{B}^{2})}$
are the single particle bulk Landau level wave functions and
${\nu}_{N X}^{\sigma}$ is the local filling factor for electrons with
spin
$\sigma$ in Landau level $N$.
For well behaved confinement potentials the partial densities can be
deduced
from the total densities $n^{\uparrow}$ and $n^{\downarrow}$. For
these
potentials the total filling factor for a given spin direction,
$\nu^{\sigma}(x)$, lying between $N_{max}-1$ and
$N_{max}$, always consists of $N_{max}-1$ completely filled Landau
levels and
a partially filled one. This means that the kinetic energy (and also
the
exchange energy, see Eqs.~(\ref{exchen}) and (\ref{genldaexch})) is
still a functional of the {\em total} and not the partial densities. 

The second contribution is the confinement energy which is given by:
\begin{equation}
E_{c}[n] = \int_{-\infty}^{\infty} dx \; n(x)\; e V_{c}(x).
\label{confen}
\end{equation}
The electrostatic Hartree energy is:
\begin{equation}
E_{H}[n] = \int_{-\infty}^{\infty} \;dx \; n(x)\; e V_{H}[n],
\label{harten}
\end{equation}
where the Hartree potential, $V_{H}[n]$, is itself a functional of
the total
density. For the strictly two dimensional electron gas (See 
Ref.~\onlinecite{lex} for the case of a quasi-two dimensional electron
gas) it is given by:
\begin{equation}
V_{H}[n]=- \frac{e}{2 \pi \epsilon l_{B}}\;\int_{-\infty}^{\infty} dx'
\;n(x')\;\ln{\left|x-x'\right|}.
\label{hartree}
\end{equation}
The fourth term denotes the Zeeman energy:
\begin{equation}
E_{Z}[n^{\uparrow},n^{\downarrow}] = \frac{1}{2} g \mu_{B} 
B_{0}\;\int_{-\infty}^{\infty}\;dx \left\{ n^{\uparrow}(x) -
n^{\downarrow}(x) 
\right\},
\label{zeemanen}
\end{equation}
where $g$ is the bare Land\'e factor.

Our only concern left is the explicit form of the last term in
Eq.~(\ref{totenergy}), which is the exchange-energy functional.
In the local density approximation (LDA) it reads:
\begin{equation}
E_{x}[n^{\uparrow},n^{\downarrow}] = \int_{-\infty}^{\infty} dx
\;\left\{ n^{\uparrow}(x)\; \epsilon_{x}(n^{\uparrow}(x)) +
n^{\downarrow}(x)
\; \epsilon_{x}(n^{\downarrow}(x)) \right\}.
\label{exchen}
\end{equation}
Here $\epsilon_{x}$ denotes the one-particle exchange energy of the
homogeneous
2DEG with ground state densities $n^{\uparrow}$ and $n^{\downarrow}$,
which is
magnetic field dependent: In the absence of a magnetic field, it is
proportional
to $\sqrt{n}$ (Ref.~\onlinecite{stern}), but in the magnetic quantum
limit
considered here, the local exchange energy per electron is:
\begin{equation}
\epsilon_{x}(n^{\sigma}(x))= - \frac{e^{2}}{4 \pi \epsilon l_{B}}\;
\sum_{N N'} \; c_{N N'} \;n^{\sigma}_{N'}(x),
\label{genldaexch}
\end{equation}
where the coefficients $c_{N N'}$ describe the exchange coupling
between Landau
levels $N$ and $N'$. They can be found by calculating the exchange
energy per
particle of an extended $N$ Landau level system.\cite{kinaret} For two
occupied
levels, {\it e.g.}, they are $c_{00}=\sqrt{2 {\pi}^{3}}$,
$c_{01}=c_{10}=\sqrt{{\pi}^{3}/2}$ and $c_{11}=\sqrt{9{\pi}^{3}/8}$.

The local density approximation is clearly only justified when the
density variations are small on a characteristic length scale, which
in our case is the magnetic length. In practice, however, it often turns
out to be a useful and accurate tool even in cases in which this
condition is not met.\cite{buukske}

Because we expect the results for higher Landau levels to be
qualitatively
the same, we will in the following restrict ourselves to the magnetic
quantum limit,
for which only the lowest Landau level is occupied. In the space of
the lowest
(spin-polarized) Landau level, the kinetic energy per particle is
constant and
may be disregarded. The validity of this approximation is discussed in
Sec.~\ref{sec:exp}. The density is in this case given by: 
\begin{equation}
n(x)=\frac{1}{2 \sqrt{\pi^3}} \sum_{\sigma} \int_{-\infty}^{\infty}
dX \;\nu^{\sigma}(X)\; e^{\;-{(x-X)}^2},
\label{convolution}
\end{equation}
where we have replaced the sum over $X$ in Eq.~(\ref{genconv}) by an
integral
and substituted the lowest Landau level wave functions. This relation
shows
that in the extreme quantum limit there exists a one to one
correspondence
between the density and the filling factor which enables us to use the
latter
as the variational function in the determination of the ground state
of the system (see Sec.~\ref{sec:resdis}).

The exchange energy per electron in the lowest Landau level ($N=N'=0$)
reduces to:
\begin{equation}
\epsilon_{x}(n^{\sigma})=-\sqrt{2 \pi^{3}}\;
\frac{e^{2}}{4 \pi \epsilon l_{B}}\; n^{\sigma}(x),
\label{ldaexch}
\end{equation}
{\it i.e.} the one-particle exchange energy for a given spin
direction is just proportional to the corresponding electron density.

On substituting Eqs.~(\ref{hartree}), (\ref{convolution}) and
(\ref{ldaexch})
in Eq.~(\ref{totenergy}) for the total energy and performing the
functional derivatives, Eqs.~(\ref{chempot}) become:
\begin{mathletters}
\label{potentials}
\begin{eqnarray}
\frac{\delta E[n^{\uparrow},n^{\downarrow}]}{\delta n^{\uparrow}(x)}
&=&
e V_{c}(x) + e V_{H}[n] + 2 \epsilon_{x}[n^{\uparrow}] +
\frac{1}{2} g \mu_{B} B_{0} = \mu, \\
\frac{\delta E[n^{\uparrow},n^{\downarrow}]}{\delta n^{\downarrow}(x)}
&=&
e V_{c}(x)+e V_{H}[n] +2 \epsilon_{x}[n^{\downarrow}] -
\frac{1}{2} g \mu_{B} B_{0} = \mu.
\end{eqnarray}
\end{mathletters}
The numerical treatment of these coupled equations and the results 
will be
presented in the next section.

\section{Calculation of the ground state properties}
\label{sec:results}

\subsection{Results}
\label{sec:resdis}

As described above we obtain the ground state occupation numbers by
numerically
minimizing the
total energy with respect to the filling factors, while keeping the 
total number
of electrons constant. We discretize the filling factor and impose
the boundary
conditions: $0\leq\nu^{\sigma}\leq1$. The numerical algorithm that 
minimizes
the total energy uses a sequential quadratic programming method and 
is very stable.
Numerical integration of the singular integrand in Eq.~(\ref{hartree})
requires some care.
By sampling the integrand equidistantly and
symmetrically around the divergence, $x=x'$, and by sampling enough 
points
to avoid oscillatory behavior in the resulting integral.
An advantage of the present numerical scheme is that it can be very 
easy
generalized to different confinement potentials. It should
be noted, however, that the choice of confinement potential is not 
completely
arbitrary. For {\it e.g.} a hard wall potential there is a stronger 
mixing
with higher Landau levels near the edges and for such a confinement
Eqs.~(\ref{convolution}) and (\ref{ldaexch}) may be no longer 
accurate. 

With the approach outlined above we can obtain, to a very good 
approximation,
both the electrostatic solution (by neglecting the exchange term) and 
the
Hartree-Fock solution as calculated by 
Dempsey~{\it et~al.}\cite{dempsey}
(by forcing the filling factors to be integer valued). Therefore we 
expect our
approach to give a good description of the intermediate regime between
these two extremes.
 
In Fig.~\ref{fig:fillfact} ground state occupation numbers obtained 
with
different methods are plotted as a function of the confinement 
strength $\alpha$.
Fig.~\ref{fig:fillfact}(a) shows the purely electrostatic solution, 
the
Hartree-Fock solution is plotted in Fig.~\ref{fig:fillfact}(b) and the
solution including exchange in the local density approximation is 
shown in
Fig.~\ref{fig:fillfact}(c). The small incompressible regions found in 
the
electrostatic solution are caused by Zeeman splitting.

The calculations were performed using a magnetic field of 7.2~T which 
for
filling factor $\nu=2$ in the bulk corresponds to a zero-field bulk 
density
of $3.5{\times}10^{11}$~cm$^{-2}$. For GaAs the static dielectric 
constant 
$\epsilon_{r}=12.5$ and the bare Land\'e factor $|g|=0.44$. The 
filling factors
are plotted for a constant number of electrons and confinement 
strengths
$\alpha=$~0.035,~0.041 and 0.047 respectively. Converted to energy 
level
spacings of a parabolic confinement potential of the form:
$\frac{1}{2} m^{*} \omega_{0}^{2} x^{2}$ these values correspond to
$\hbar \omega_{0}=3.4$~meV, 3.8~meV and 4.1~meV respectively.

As can be seen from Fig.~\ref{fig:fillfact}(c) the effect of exchange 
on the 
electrostatic solution is a reduction of the compressible regions
because it favors integer filling. However, a comparison with 
Fig.~\ref{fig:fillfact}(b)
shows that for soft confinement potentials the Hartree-Fock 
approximation no
longer holds, {\it i.e.} the electrostatic interaction overcomes the
tendency of exchange to form an incompressible ground state. Note, 
however,
that the solution for $\alpha = 0.047$ in
Fig.~\ref{fig:fillfact}(c) is almost identical to the corresponding 
Hartree-Fock
solution shown in Fig.~\ref{fig:fillfact}(b), which shows that for 
confinement
potentials that are strong enough the solution is forced into the 
integer
filling regime, in agreement with the qualitative picture given in
Ref.~\onlinecite{dempsey}. Note that this transition from 
incompressible to
compressible state is a genuine correlation effect since it 
corresponds to the
mixing of different single-determinant configurations, in spite of our
exchange-only potential.

It is clear from Fig.~\ref{fig:fillfact}(c) that
the width of the incompressible region is of the same order of 
magnitude as the
width of the outermost compressible strip, even for the relatively 
wide strips
we consider here (approximately 420~nm for a magnetic field of 7.2~T).
This 
in contrast to the electrostatic description of edge states shown in 
Fig.~\ref{fig:fillfact}(a), where the width of the incompressible 
region is
always much smaller than that of the compressible one. However, for 
soft
confinement potentials the solution including exchange does not 
deviate
dramatically from the purely electrostatic one, which obviously 
implies that
the Hartree-Fock approach breaks down in this regime. Only a strong 
confinement
potential can cause incompressible edge channels, although in
Fig.~\ref{fig:fillfact}(c) the confinement is still not strong enough 
to
reduce the splitting to that of the bare Zeeman splitting. In the case
of even
harder confinement
($\alpha \approx 0.073$,~{\it i.e.}~$\hbar \omega_{0} \approx 
5.3$~meV) we find
that the splitting indeed reduces to this minimum value (not shown).

Note that the entire range from the electrostatic to the (unpolarized)
Hartree-Fock regime can
be realized in realistic confinement potentials with 
$\hbar \omega_{0}$
ranging from 3.4~meV to 5.3~meV. Confinement potentials with level 
spacings of a
few meV have been realized\cite{mceuen,ashoori} so it should be 
possible to
test these results by experiments. A possible method to stiffen the 
confinement
experimentally is fabricating the 2DEG closer to the surface of the 
AlGaAs/GaAs
heterostructure.

In essence, our method to include exchange is using an position 
dependent
$g$-factor that depends on the local density. A simple approximation
to this method would be to use the enhanced $g$-factor corresponding 
to a
single occupied spin level.\cite{janak} This method does increase the 
width
of the incompressible strip but in general does not reproduce the 
qualitative
and quantitative features obtained with our method.\cite{theo}

In Fig.~\ref{fig:potentials} the calculated one-particle potentials 
in the
wire are plotted for both spin directions. These potentials correspond
 to the 
ground state density distribution of Fig.~\ref{fig:fillfact}(c) with
$\alpha = 0.035$ which is also included in the figure.
Fig.~\ref{fig:potentials}(a) shows the electrostatic potential, which 
consists
of the Hartree and confinement potential. Also plotted are the 
exchange
potential and the total potential for the majority-spin electrons, 
{\it i.e.}
electrons occupying the lowest spin level.
In Fig.~\ref{fig:potentials}(b) the same can be seen for the 
minority-spin
electrons. In the figures the constant Zeeman term has been omitted 
for
simplicity so that the total potential is just the sum of the 
electrostatic
and exchange potential.

Fig.~\ref{fig:potentials}(a) shows that for a majority spin electron 
the
energetically most favourable position is near the edges. This is 
because
these electrons are forced by exchange to form an electrostatically
unfavourable density profile in which no screening of the confinement 
potential
is possible, {\it i.e.} the density of majority
spin electrons is constant in the bulk of the wire. This in contrast
to the minority spin electrons, for which the potential in the bulk is
 flat. 
This is due to the fact that these electrons still have the freedom to
 form a
density distribution which can screen the external potential without 
being
influenced much by the exchange interaction.

The typical voltage drop from the edge to the middle of the wire has
experimentally realistic values of $\approx$~300~mV for the bare 
confinement
potential and $\approx$~30~mV for the selfconsistent potential.

We also performed calculations of the widths of compressible and 
incompressible
regions as a function of the confinement strength.
The results are visible in Fig.~\ref{fig:phasetr} where the widths of 
the 
outermost incompressible region (I) and innermost compressible region
(C) have been plotted as a function of $\alpha$. If no incompressible
strip is present in the middle of the wire, C is defined as half the 
width of
the total compressible middle region (see inset). In the figure the 
points do
not always lie on the smooth curve which serves as a guide for the 
eye. This is
due to the spacing between sampling points (0.2~$l_{B}$), which 
imposes a upper
bound on the accuracy of the calculated widths. 

The compressible strip shrinks rapidly as the confinement is 
increased but if the width has reached a value of roughly a few 
magnetic
lengths, the further decrease becomes very
slow and in practice we always find a small but finite compressible 
region.
It is clear, however, that a Hartree-Fock treatment should give good 
results
for hard confinements where small compressible regions exist. We 
recover
the spin-polarizing transition of Ref.~\onlinecite{dempsey}
since the incompressible strip between spin-up and spin-down
channels (I in Fig.~\ref{fig:phasetr}) drops to zero at a certain 
critical
value of the confinement strength (neglecting Zeeman splitting). Both 
the value
for the critical confinement strength and the overall shape of the 
curve agree
well with the results of Ref.~\onlinecite{dempsey}.

An important consequence of exchange on the distribution of electrons 
for
experiments is the increased separation between the compressible 
regions.
This causes a strong decrease in inter-edge channel scattering because
of the
reduced overlap of the edge-channel wave functions, which leads to an
increase
in the spin-flip equilibration length of the edge channels. This in 
contrast to
the assumption of an edge channel separation of the order of one 
magnetic length
in Ref.~\onlinecite{khaetskii}.
Using a typical edge channel separation of a few magnetic lengths in 
accordance with the present results, Khaetskii's 
theory\cite{khaetskii} would
give a much longer equilibration length.

The exchange enhanced channel separation has also consequences for
the two-terminal magnetoconductance of a narrow channel or point 
contact. 
In Ref.~\onlinecite{cms}, Chklovskii~{\it et al.} discuss the 
conductance
for these systems in the framework of their electrostatic description
of
the channels. They obtain conductance quantization but, due to the 
small width
of the incompressible regions, the calculated plateau widths are much
smaller than those experimentally observed. They attribute this 
discrepancy 
between theory and experiment to the presence of disorder. However, 
we propose
as an alternative explanation the electronic exchange interaction 
which we find
to strongly enhance the width of the insulating strips.

\subsection{Accuracy}
\label{sec:exp}

In this subsection we study the accuracy of our theory
and how our present results are altered if we take into account 
realistic
features which do not change the qualitative results but are important
to
describe experimental situations.

First we investigate the validity of Eq.~(\ref{convolution}) for high
magnetic fields where mixing with higher Landau levels was 
disregarded.
To this end we use a parabolic confinement
$\frac{1}{2} m^{*} \omega_{0}^{2} x^{2}$
and a density distribution of the form:
$n(x)=\frac{1}{2 \pi l^{2}} \theta(x-W/2) \theta(x+W/2)$, where $W$ 
is the
width of the wire.
The 2DEG in the presence of the uniform magnetic field and parabolic
confinement potential is our unperturbed system and we treat the 
Hartree
potential as a perturbation. The energy levels of the unperturbed 
system are
$E_{nk}=(n+\frac{1}{2}) \hbar \Omega + \hbar^{2} k_{y}^{2}/2M$,
where $\Omega = \sqrt{\omega_{c}^{2} + \omega_{0}^{2}}$ is the 
renormalized
cyclotron frequency and $M$ is the renormalized electron mass:
$M=m^{*}(1+{\omega_{c}^{2}}/{\omega_{0}^{2}})$. The wave functions 
are the
bulk Landau level wave functions with effective magnetic length 
$l=l_{B} 
\sqrt{{\omega_{c}^{2}}/{(\omega_{c}^{2}+
\omega_{0}^{2})}}$.\cite{kinaret} 
According to first order perturbation theory, the ground state wave 
function, 
$ \left| \psi \right>$, of the electron at the edge, with quantum 
number
$X=W/2$, is approximately:
\begin{equation}
\left| \psi \right> = \left| \psi_{0} \right> + 
\frac{\left< \psi_{1} \right| e V_{H} \left| \psi_{0} 
\right>}{E_{0}-E_{1}} 
\left| \psi_{1} \right>,
\label{perturb}
\end{equation}
where we have used the Hartree potential as the perturbation.
A straightforward calculation of the coefficient of 
$\left| \psi_{1} \right>$
(which of course depends on $W$) shows that for 
$20~l_{B} \leq W \leq 40~l_{B}$,
a magnetic field of 5~T and a typical value of 
$\hbar \omega_{0}=3$~meV for the
parabolic confinement, the mixing with the first Landau level wave 
function is
approximately 10\% and that this value decreases as 
$1/\sqrt{B_{0}}$ with
increasing magnetic field. We have to conclude that the dip in the
electrostatic, {\it i.e.} Hartree and confinement, potential at the 
edges
which is responsible for the spin-polarizing transition and ultimately
 for the
compressible edge state, cannot be compensated by mixing with higher 
Landau
levels, {\it i.e.} that our approximation is allowed in the high 
magnetic
field limit considered here. By performing a similar calculation it 
is
possible to show that the mixing due the exchange-correlation 
potential used
by Ferconi and Vignale\cite{ferconi} is only a few percent.
 
Another approximation which gives rise to quantitative deviations from
 our
theory is the assumption that the electron gas is strictly two 
dimensional.
In reality the electron wave functions also extend in the 
$z$-direction and a
form factor $F(q)$ in the Coulomb interaction,
$V(q) = e^{2} F(q)/ 2 \epsilon |q|$, takes this effect into account.
Calculations of edge channel splitting including a form factor in the
Hartree-Fock approximation have already been calculated by Rijkels and
Bauer.\cite{lex} By comparison with Ref.~\onlinecite{lex}
we estimate the values of $\alpha$ to decrease by about
0.004 if a form factor would be included into our calculations.
This shifts the range of $\hbar \omega_{0}$ in which both the 
electrostatic and
the Hartree-Fock regime are unreliable to lower energies, {\it i.e.} 
to
3.2~meV-5.1~meV.

\section{Applications}
\label{sec:applications}

The present theory can be easily generalized to a variety of systems. 
In the
next subsections we consider two special cases, which are 
experimentally
relevant. 

\subsection{Back gate}
\label{sec:backgate}

As a first application we have carried out calculations for a system 
like
the one considered in Sec.~\ref{sec:system}, but now in the presence 
of an
infinite plane of constant potential at a distance $d$ from the 2DEG, 
which may
represent a real back gate or a plane of not fully depleted donors 
parallel to
the 2DEG. This gate can be easily included into the calculations by 
adding the
potential of a mirror charge distribution of opposite sign located at 
a distance
$2d$ from the electron gas:
\begin{equation}
V_{bg}(x)=\frac{e}{2 \pi \epsilon l_{B}}\;\int_{-\infty}^{\infty}\;
dx'\;n(x')\; \ln{\sqrt{{(x-x')}^{2}+{(2d)}^{2}}}.
\label{backgatepot}
\end{equation}
In Fig.~\ref{fig:nubackgate} we have plotted the solutions that 
minimize the
total energy of the system including the back gate as a function of 
the
distance $d$ to the wire for a given number of electrons and a fixed
confinement potential.

It is clear from the figure that moving the back gate closer to the 
2DEG has
the effect of reducing the width of the compressible regions. This is 
to be
expected because the back gate screens the long-range Coulomb 
interaction in
the electron gas, thus effectively reducing the importance of the 
direct
interaction relative to the exchange interaction and thereby forcing 
the
solution for small distances $d$ into the Hartree-Fock regime.

It should be noted that the 
description of the back gate given here does not apply for a quantum 
wire
defined by a depletion gate. In that case the back gate would also 
screen
the charge on the depletion gates, thus increasingly altering the 
confinement
potential as the back gate is moved closer to the 2DEG. In that case 
a
fully self-consistent calculation is necessary. Furthermore, for our 
results
to be correct, the distance from the gate to the 2DEG should not be 
too small
since in that case also the short-range exchange interaction would be 
influenced
by the gate.

\subsection{Ground state of a quantum dot}
\label{sec:qdot}

A system that is very similar to the strip previously considered is a 
quantum
dot in a high magnetic field, especially when the number of electrons 
in the
dot is large. There are two main differences. In the first place we 
use the
symmetric gauge to describe the dot. In this gauge the wave functions 
in the 
lowest Landau level, labeled by the quantum number of angular 
momentum, are 
${\phi}_{m}(z)=\frac{1}{\sqrt{2 \pi}} \frac{z^{m}}{\sqrt{m!}} 
e^{-|z|^{2}/2}$
where $z=\frac{x+iy}{\sqrt{2}}$. The second difference is that 
because the
system is finite the filling factor $\nu$ is no longer labeled by the 
continuous
variable $X$ but by the discrete quantum number $m$.
As a consequence the radial symmetric density is now given by:
\begin{equation}
n(r) = \frac{1}{2 \pi} \sum_{m \sigma} \frac{\nu_{m}^{\sigma}}{m!}\;
\exp{\left( -\frac{r^{2}}{2} \right)} \; {\left( \frac{r^{2}}{2} 
\right)}^{m},
\label{densqdot}
\end{equation}
where $\nu_{m}^{\sigma}$ is the filling factor for a given angular 
momentum
$m$ and spin direction $\sigma$. We impose as boundary conditions that
$0\leq\nu_{m}^{\sigma}\leq1$ and that the total charge in one spin
level $\sum_{m} \nu_{m}^{\sigma}$ is an integer.
Due to the circular symmetry of the dot the Hartree potential is in 
this case
given by:
\begin{equation}
V_{H}(r) = -\frac{e}{4 \pi \epsilon l_{B}} \int_{0}^{2 \pi} 
d \phi \int_{0}^{\infty} dr' \frac{r' n(r')}{\sqrt{r^{2} + 
{r'}^{2} - 2 r r'
\cos{\phi}}}.
\label{hartreeqdot}
\end{equation}

An analysis analogous to the one presented in
Sec.~\ref{sec:exp} shows that for high magnetic fields (approximately 
5~T) the
mixing with higher Landau levels due to the Hartree potential and the
exchange-correlation potentials used in Ref.~\onlinecite{ferconi} is 
only a few
percent. As a consequence, fractional occupation numbers are 
essential, 
which cannot be described by the Kohn-Sham scheme of 
Ref.~\onlinecite{ferconi}.

We have calculated the ground state of a dot consisting of 40 
electrons
using a parabolic confinement potential
$V_{c}(r) = \alpha\;\frac{e}{4 \pi \epsilon l_{B}}\;r^{2}$ with 
varying
strength $\alpha$ in a magnetic field of 5.0~T. To find the ground 
state for a
given magnetic field we vary the distribution of electrons among the 
occupied
spin levels while keeping the total number of electrons constant. 
The results are plotted in Fig.~\ref{fig:nuqdot}. The ground states 
are similar
to those of the strip except for the fact that the middle region is 
wider.
This is due to the fact that the wave functions are centered around 
the radii
$r_{m} = \sqrt{2 m}~l_{B}$, where $m$ is the angular momentum index, 
and not
around equidistant points like in the strip.

In Ref.~\onlinecite{mceuen} a selfconsistent calculation was 
performed for the
density profile of a quantum dot containing 39 electrons using a
modified Hartree form for the electrostatic electron-electron 
interaction:
\begin{equation}
V_{ee}({\bf r},{\bf r'}) = -\frac{e}{4 \pi \epsilon l}
\left\{
\frac{1}{\sqrt{ {|{\bf r}-{\bf r'}|}^{2} + {\langle \delta z 
\rangle}^{2} }} -
\frac{1}{\sqrt{ {|{\bf r}-{\bf r'}|}^{2} + {4 d}^{2} }}
\right\},
\label{veemceuen}
\end{equation}
thus taking into account the finite $z$-extent $\langle \delta z 
\rangle$ of
the wave functions and a back gate at distance $d$. 
Again we want to extend these calculations to include exchange but we 
must keep
in mind that a finite $\langle \delta z \rangle$ also reduces the 
exchange
interaction by a form factor:
\begin{equation}
F(\langle \delta z \rangle) = \exp{\left( \frac{1}{2}
{\langle \delta z \rangle}^{2}\right )}
\mbox{Erfc} \left( \frac{\langle \delta z \rangle}{\sqrt{2}} \right),
\label{formfactor}
\end{equation}
where $\langle \delta z \rangle$ is given in units of the magnetic 
length and
$\mbox{Erfc}($x$)=1-\mbox{Erf}(x)$ denotes the complementary error 
function.
A calculation using interaction Eq.~(\ref{veemceuen}) and the local
exchange contribution reduced by the form factor 
Eq.~(\ref{formfactor}) gives
a ground state which differs from that of McEuen~{\it et al.} in the 
sense that
the widths of the incompressible (compressible) regions are larger 
(smaller) in
our solution.

We have calculated the addition spectrum of dots consisting of 37 and 
38
electrons in order to determine the effect of exchange on the 
inter-level
transitions that are responsible for the oscillatory behavior of the 
addition
energy.
Here we take into account the fact that the magnetic
length is renormalised by the magnetic field, Eq.~(\ref{renmaglen}),
in order to compare the energies at different magnetic fields.
The addition energy $\Delta E(N)$ needed for adding an extra electron 
to a dot
of $N$ electrons is:
\begin{equation}
\Delta E(N) = E(N+1)-E(N),
\label{addenergy}
\end{equation}
where $E(N)$ is the total energy given by Eq.~(\ref{totenergy}). The 
results
are depicted in Fig.~\ref{fig:addition}. It should be mentioned that 
our
results are very close to the Hartree-Fock solutions. This is due to 
the fact
that the combination of the back gate and exchange force the solution 
into the
integer filling regime.

The inset of Fig.~\ref{fig:addition} shows the energy dependence of 
different
dot configurations with magnetic field for a dot containing 39 
electrons.
The ground state is formed by the configuration with lowest energy.
The numbers in the inset represent the number of electrons in the 
majority
({\it i.e.} lowest) spin level for the indicated curve. As the 
magnetic field
increases, the degeneracy of the spin levels is increased and at 
certain
magnetic fields an electron can jump from the upper to the lower 
level, 
causing the kinks in the ground state energy. When all electrons 
are
in the lowest spin level no transitions can occur and the
curves for both the total and addition energy are smooth.

Recent calculations\cite{chamon} and measurements\cite{klein}
indicate that at higher magnetic fields than the ones shown in
Fig.~\ref{fig:addition} an edge reconstruction occurs, resulting
in an extra kink in the curve for the addition energy in the regime
where only one spin level is occupied. Performing a calculation
with filling factors that are restricted to integer values, we are
able to reproduce this feature. However, for the small confinement
strength $\alpha$ at which this transition occurs, the
fractional-filling
solution is not so well converged. In spite of the fact that we find
evidence for the edge reconstruction, the numerical accuracy of our
solution does not allow decisive conlusions.

In comparison with the results obtained in Ref.~\onlinecite{mceuen} 
the
oscillations found in our calculations have a larger period and 
amplitude by
a factor of approximately 1.5. The increase in amplitude of the 
oscillations
can be qualitatively understood in terms of the capacitance model 
for the
island proposed by Evans~{\it et al.}\cite{evans} and is due to the 
increase
of the width of the incompressible strip in the picture including 
exchange.
This results in a decrease of the capacitance between the inner and 
outer
compresible regions where the extra electron can be added and hence 
in an
increase in the peak-to-peak amplitude. The larger period can be 
explained in
a similar fashion: Exchange effects reduce the widths of the 
compressible strips
and thus the capacitances $C_{1}$ and $C_{2}$ between the
respective compressible strips and the back gate which results in an 
increased
period of the oscillations compared to the electrostatic treatment of
McEuen~{\it et al.}.
Compared with their experimental data, the period of the oscillations 
found
in our calculations is about 0.3~T too large whereas their 
calculations
give a period which is 0.25~T too small. However, these values depend
sensitively on the calculated ground state density since the total 
energy,
with a typical value around 1~eV, has to be calculated with an 
absolute accuracy
of at least 0.1~meV in order to resolve the oscillations and therefore
small numerical deviations from the true ground state would
drastically influence the addition energies.

\section{Summary and concluding remarks}
\label{sec:conclusions}

We have used density-functional theory in strong magnetic fields to 
investigate
the effects of the electron-electron interaction on edge states in 
the integer
quantum Hall regime. We have included exchange in the local density
approximation. In this approximation smooth density distributions 
corresponding
to fractional filling at the edges can be treated, which is beyond a
Hartree-Fock treatment of edge channels.\cite{dempsey}
To describe fractional occupation could otherwise only be achieved by 
exact
diagonalization or configuration interaction 
calculations.\cite{chamon}
We have found that 
the width of the edge channels is strongly reduced due to exchange 
effects and
that the width of the incompressible strips is of the order of several
 magnetic
lengths, which should have a strong influence on the spin-flip 
equilibration
length and the two-terminal conductance of a point contact. We predict
 a range
of confinement potentials $\hbar \omega_{0}$=3.2~meV-5.1~meV in which 
the
entire regime from electrostatic to Hartree-Fock is covered.
Furthermore we have calculated ground state properties of the system 
in the
presence of a back gate parallel to the quantum wire. We established 
an
increasingly important role for the exchange interaction as the back 
gate is
closed in on the 2DEG.
As a second application we have calculated the ground state of a 
quantum
dot in a strong magnetic field. We found that the results are similar
to those found for the quantum wire. We compared addition energies
for the dot with the selfconsistent calculations of 
McEuen~{\it et al.}\cite{mceuen} and found that the oscillations in 
the
addition energies have larger amplitudes and periods in our 
calculations.

\acknowledgments
The authors would like to thank Sasha Khaetskii, Yuli Nazarov and Lex 
Rijkels
for valuable discussions and A.H. MacDonald for a reprint of
Ref.~\onlinecite{mej}. This work is part of the research program of 
the
"Stichting voor Fundamenteel Onderzoek der Materie" (FOM), which is 
financially
supported by the "Nederlandse Organisatie voor Wetenschappelijk 
Onderzoek"
(NWO).

\begin{figure}[p]
\caption{Schematic picture of the system, consisting of a strip of 
2DEG in a
uniform magnetic field. The wire is confined by a parabolic potential 
in the
$x$-direction. For details see text.}
\label{fig:system}
\end{figure}

\begin{figure}[p]
\caption{Electronic ground state filling factors, (a) ignoring 
exchange and
including exchange in (b) the Hartree-Fock approximation and (c) the 
local
density approximation. The filling factors are plotted for confinement
strengths $\alpha=0.047$ (solid line), $0.041$ (dashed line) and 
$0.035$
(dashed-dotted line). Calculated for a magnetic field of 7.2~T.}
\label{fig:fillfact}
\end{figure}

\begin{figure}[p]
\caption{One-particle potentials and density distribution for the 
ground state
of Fig.~2(c) with $\alpha = 0.035$ for (a) majority-spin electrons 
and 
(b) minority-spin electrons.
Plotted are the electrostatic potential consisting of the confinement 
and the
Hartree potential (dashed line), the exchange potential (dashed-dotted
 line)
and the total potential excluding the Zeeman term (solid line). 
In the regions
where the exchange potential is zero the total and Hartree potential 
coincide.
The potentials are offset for clarity and are given in units of
$\frac{e}{4 \pi \epsilon l_{B}}$.}
\label{fig:potentials}
\end{figure}

\begin{figure}[p]
\caption{Width of innermost compressible region (C) and outermost 
incompressible
region (I) (see inset), plotted as a function of the confinement 
strength
$\alpha$. The solid lines are drawn as a guide for the eye.}
\label{fig:phasetr}
\end{figure}

\begin{figure}[p]
\caption{Occupation of the system in the presence of the back gate at 
a
distance $d=\infty$ (dashed-dotted line), $d=20l_{B}$ (dashed line) 
and
$d=10l_{B}$. The confinement strength $\alpha = 0.035$ and 
$B = 7.2~T$.
(solid line).}
\label{fig:nubackgate}
\end{figure}

\begin{figure}[p]
\caption{Occupation numbers of a quantum dot consisting of 40 
electrons for
confinement strengths $\alpha=0.09$ (solid line), $0.08$ (dashed line)
 and
$0.07$ (dashed-dotted line). The discrete filling factors are 
positioned at
the radii $r_{m}$ and have been connected by lines for clarity.}
\label{fig:nuqdot}
\end{figure}

\begin{figure}[p]
\caption{Addition energy for quantum dots of 37 and 38 electrons as 
a function
of magnetic field. The calculation was performed for a confinement 
strength
$\hbar \omega_{0}$ = 1.6~meV, $d$ = 100~nm and 
$\langle \delta z \rangle$ =
10~nm. The inset shows the ground state energies of the different 
possible
configurations of the dot as a function of magnetic field. The 
numbers represent
the number of electrons in the lowest spin level. Only odd numbers 
are
indicated.}
\label{fig:addition}
\end{figure}


\begin{references}
\bibitem{theo} Calculating the ground state in this manner does 
increase the
width of the incompressible region and it slightly reduces the width 
of the
edge channels but the effects are smaller than in those obtained with 
our
method.
\end{references}
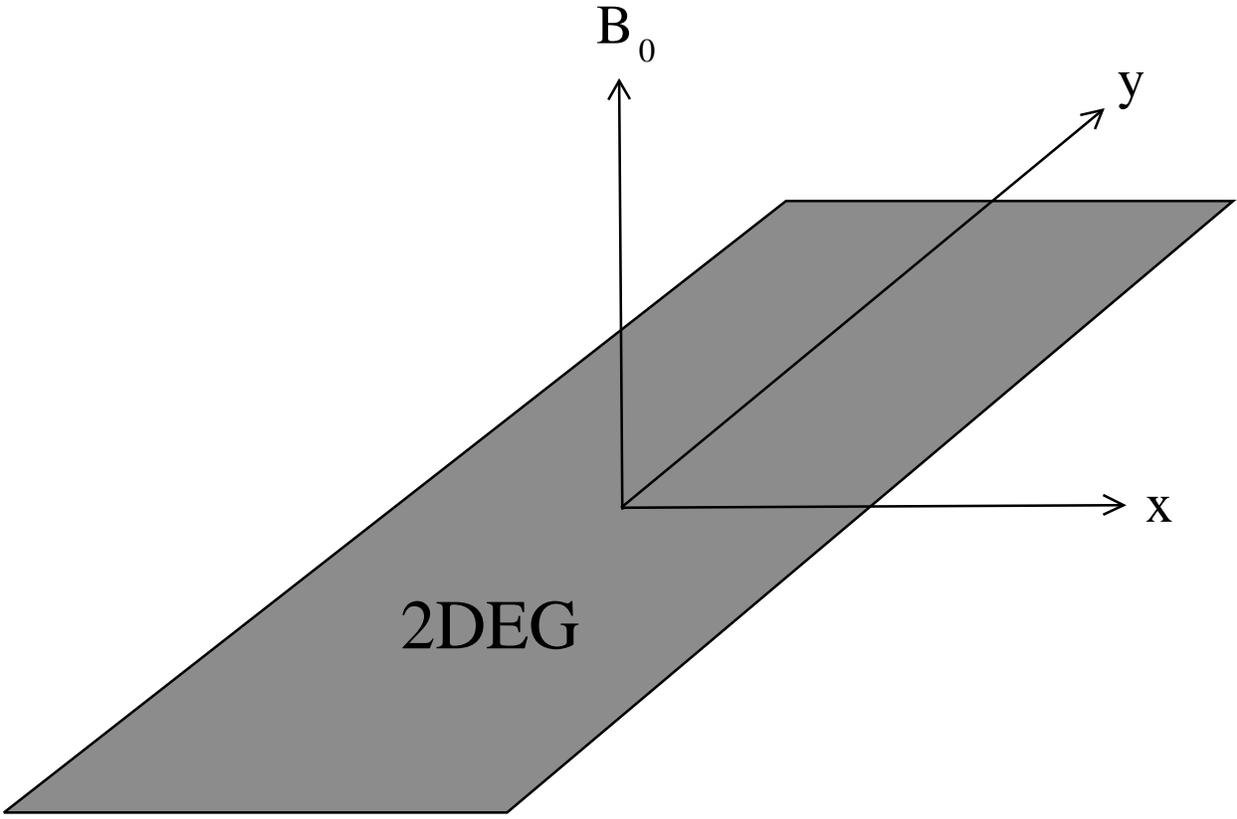
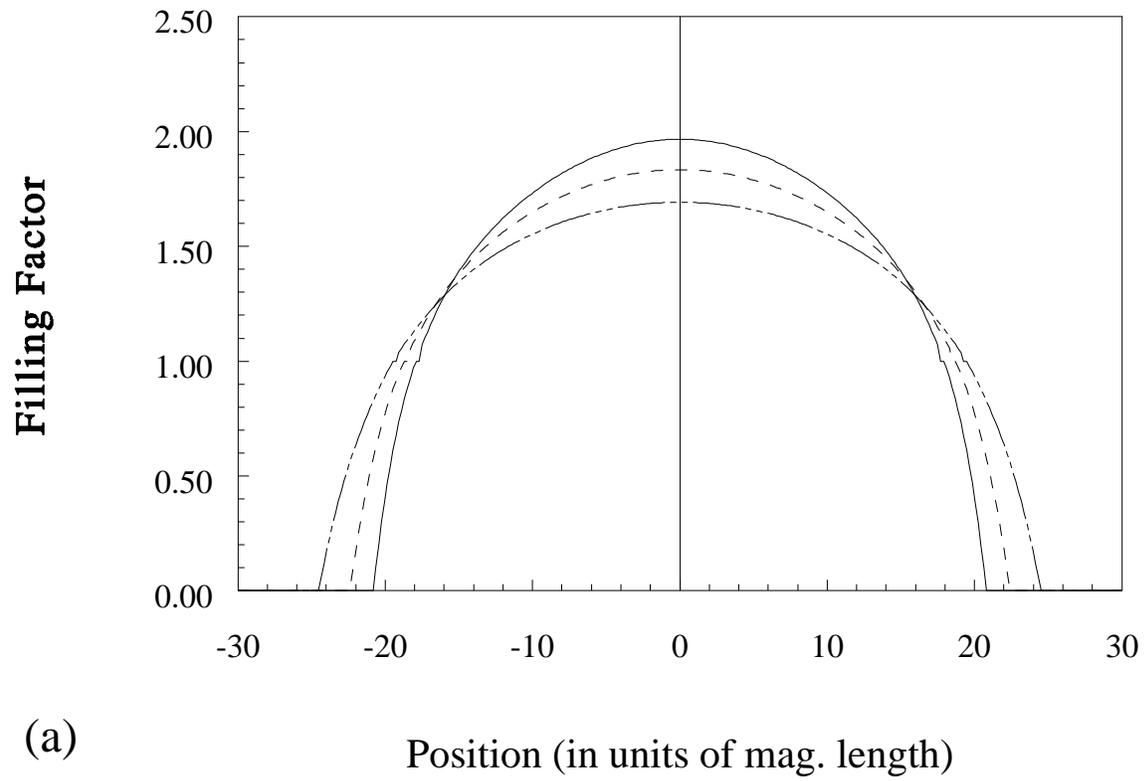
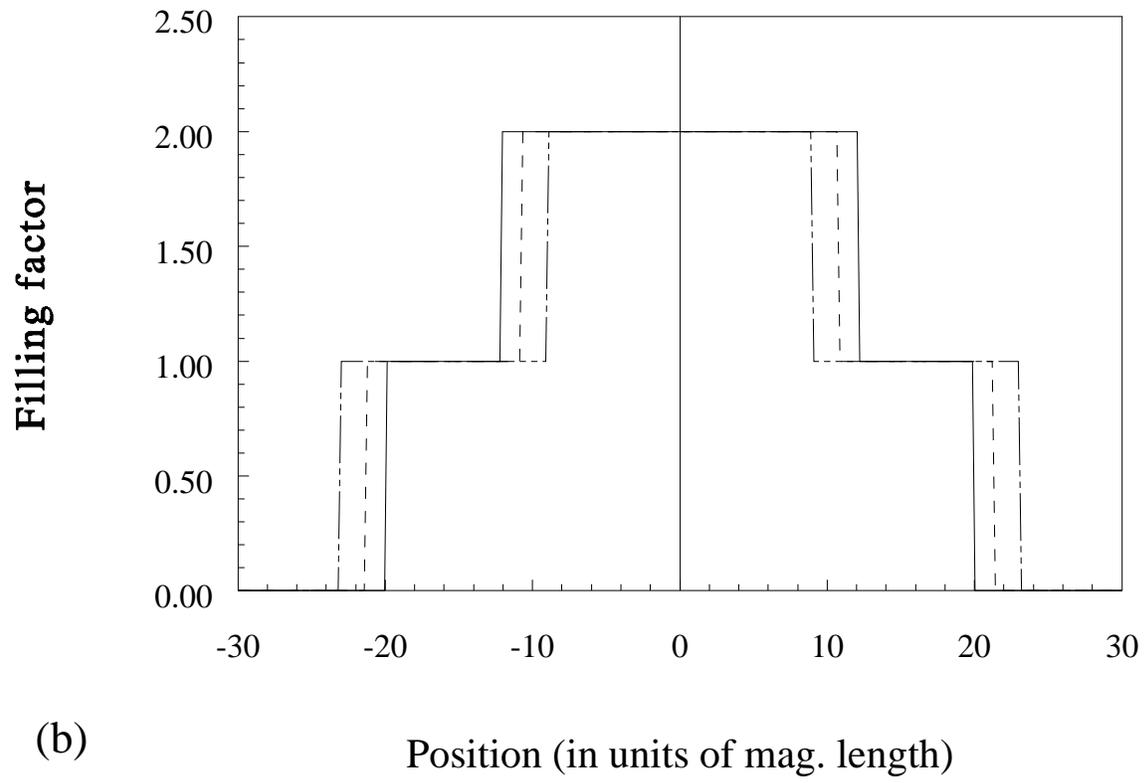
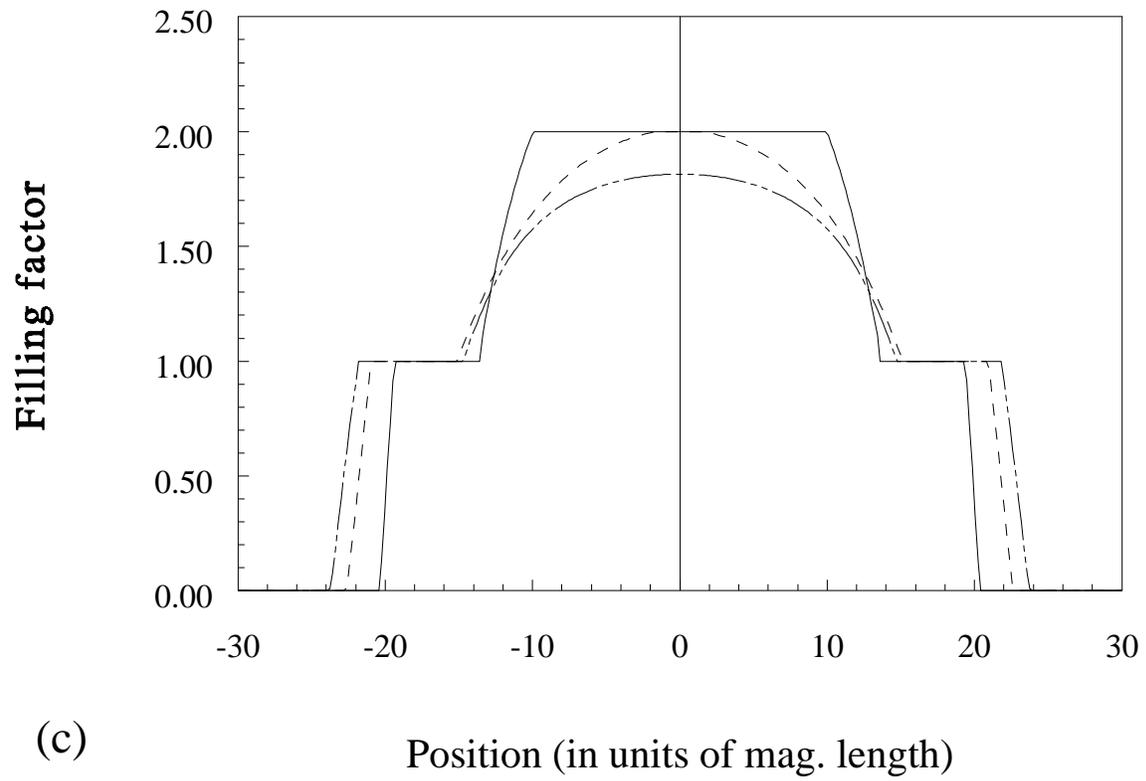
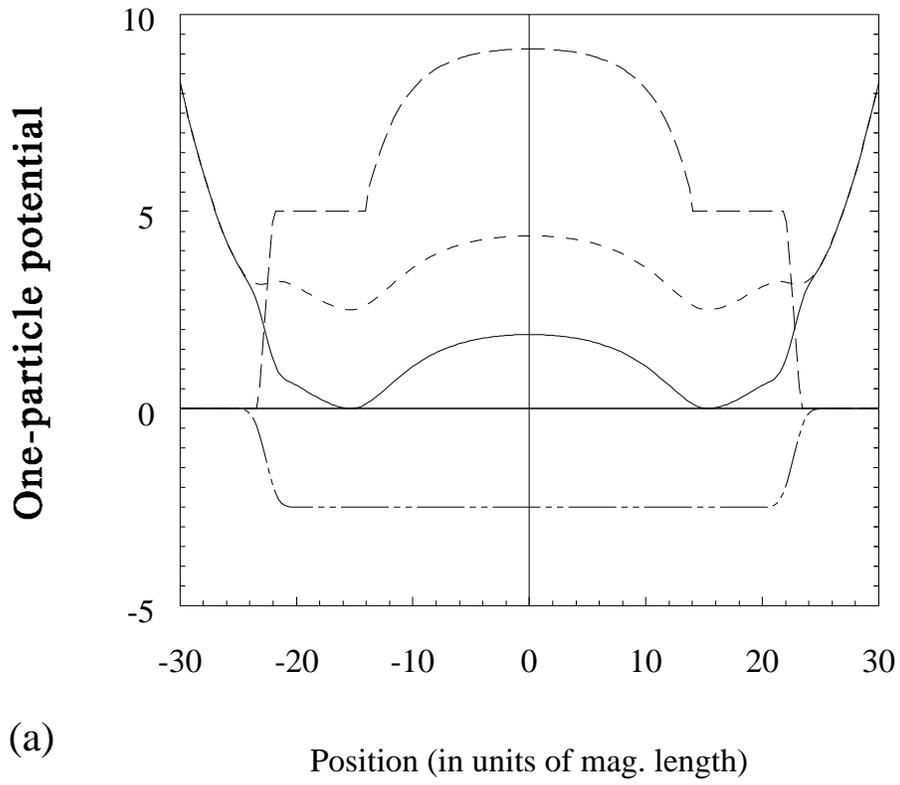
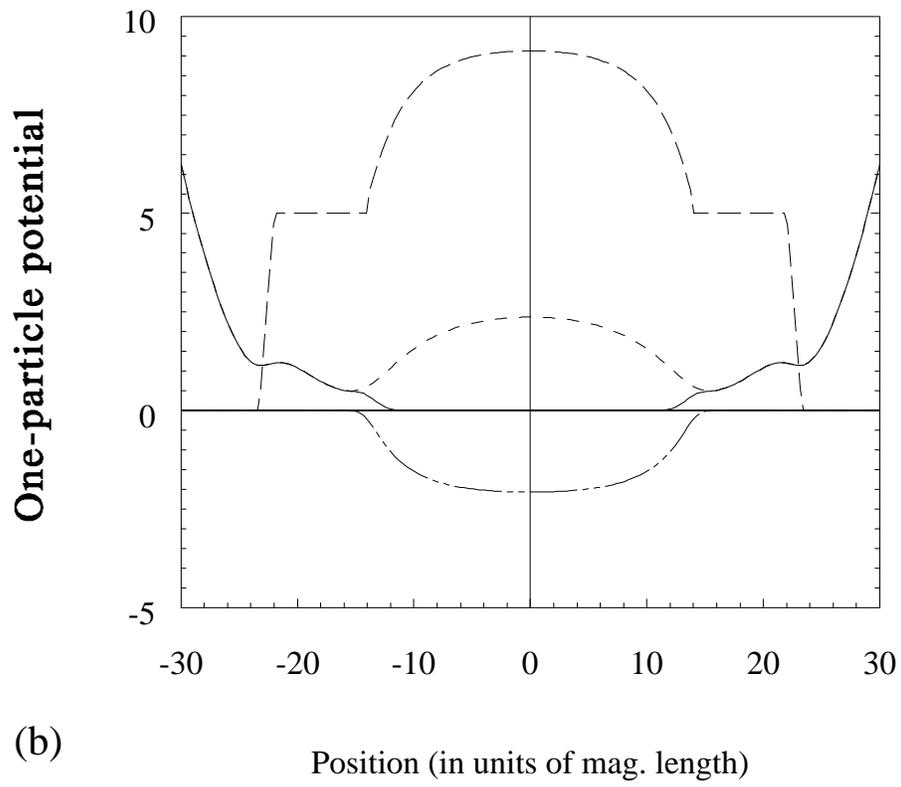
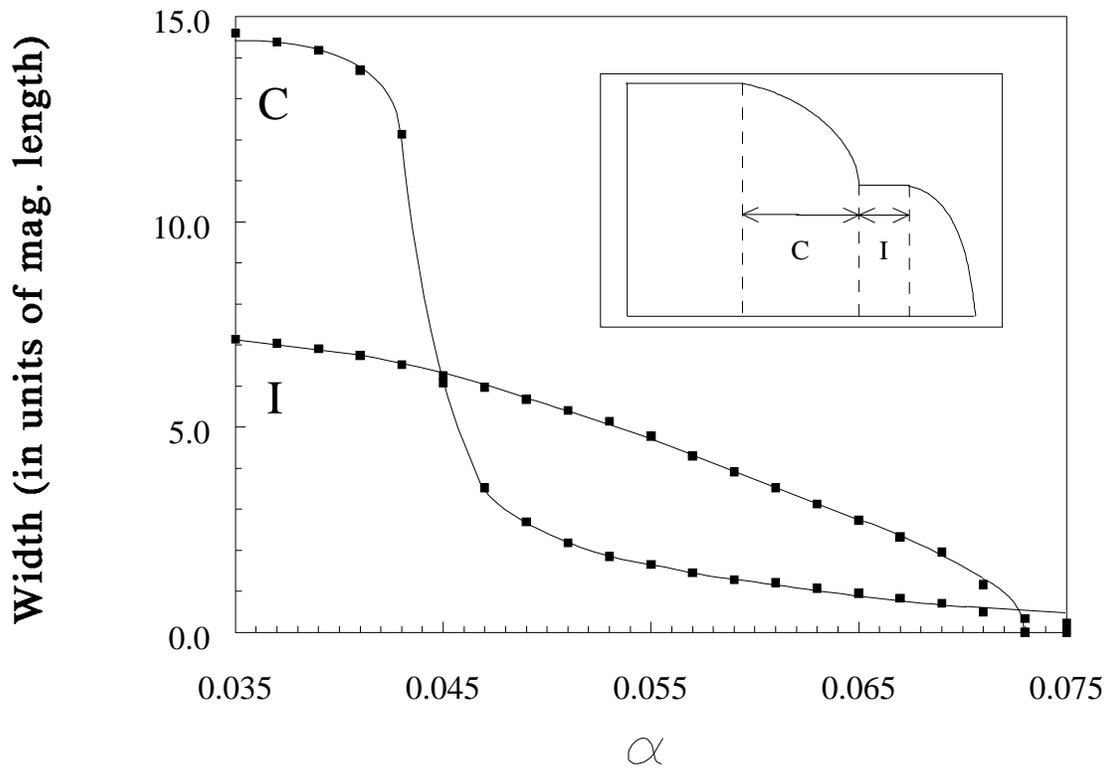
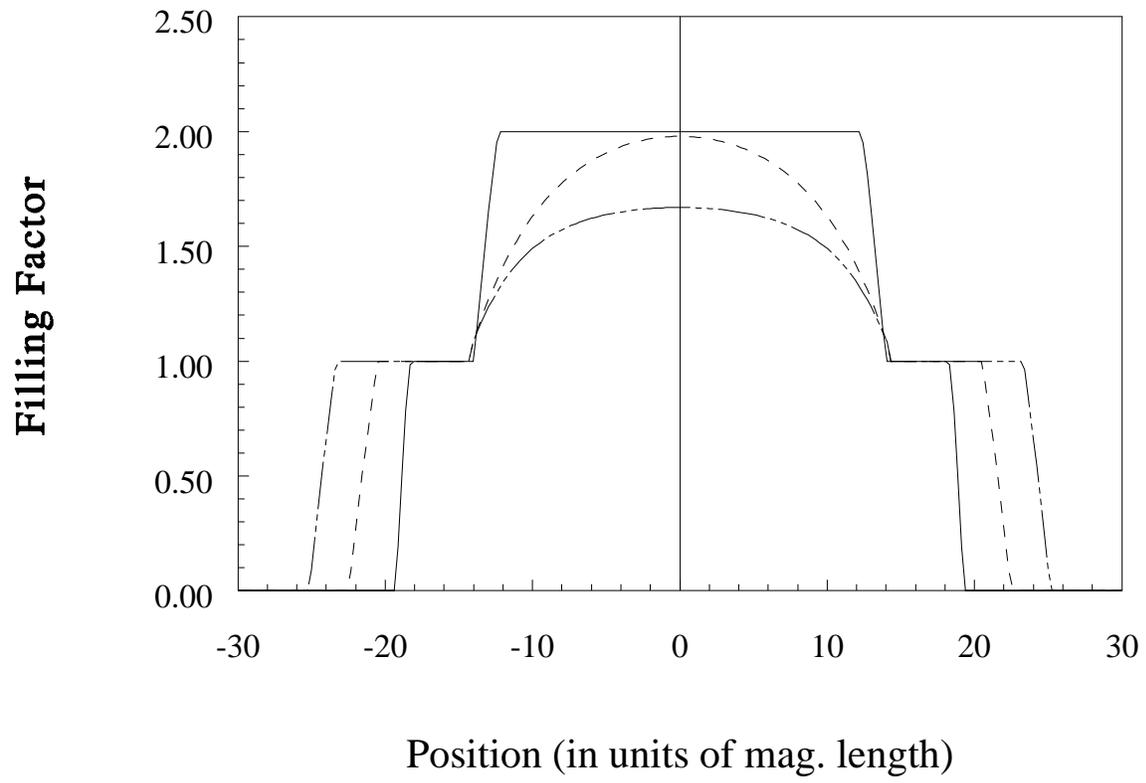
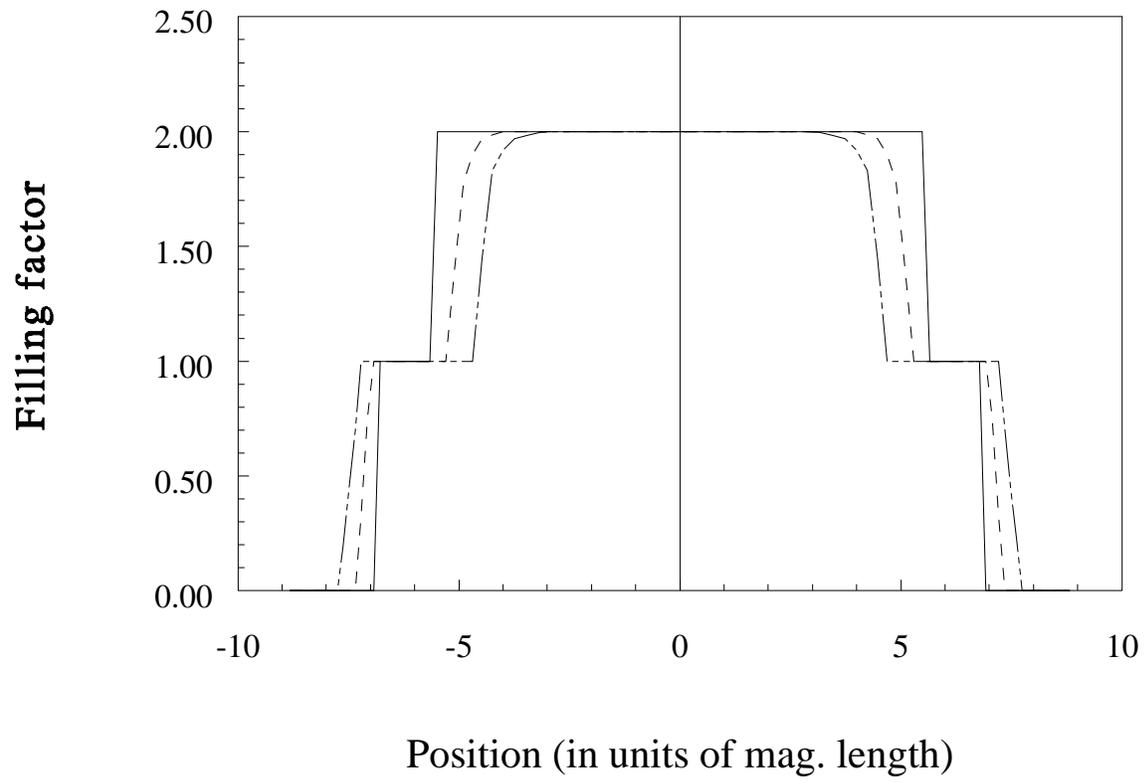
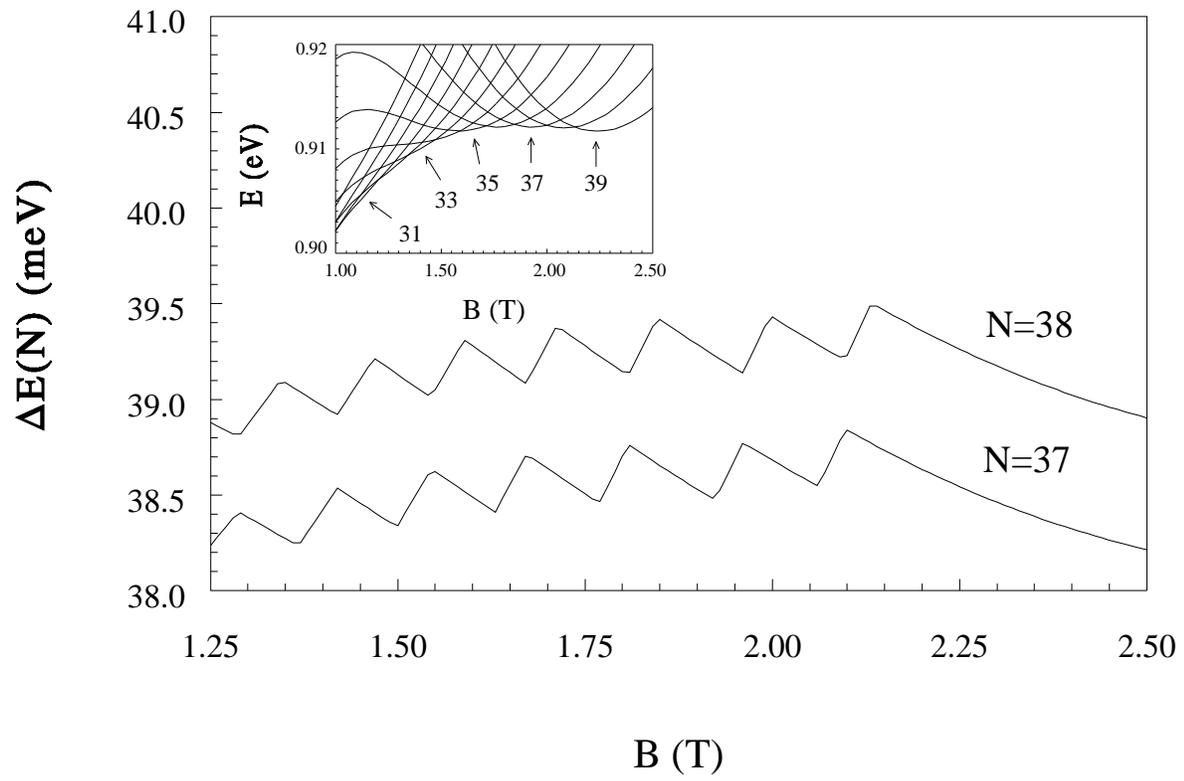
\end{document}